\documentclass[conference]{IEEEtran}
\usepackage{nomencl}
\makenomenclature

\usepackage{enumerate}
\usepackage{enumitem}
\usepackage{graphicx}
\usepackage{cleveref}
\usepackage{float}
\begin{document}
\title{A Review of Pair-wise Testing}
\author{
\IEEEauthorblockN{Jimi Sanchez}
\IEEEauthorblockA{College of Engineering and Technology\\
East Carolina University\\
Greenville, North Carolina 27858-4353\\
Email: sanchezji14@students.ecu.edu}\\   
}
\maketitle
\begin{abstract}
In software testing, the large size of the input domain makes exhaustively testing the inputs a daunting and often impossible task. Pair-wise testing is a popular approach to combinatorial testing problems. This paper reviews Pair-wise testing and its history, strengths, weaknesses, and tools for generating test cases. 
\end{abstract}
\begin{IEEEkeywords}
fault detection, Pair-wise, software testing \\
\nomenclature{$ACTS$}{Automated Combinatorial Testing For Software}
\nomenclature{$AT\&T$}{American Telephone and Telegraph Company}
\nomenclature{$Black\-box$}{Method of software testing that examines the functionality of an application without knowledge of how the internal workings are implemented.}
\nomenclature{$CATS$}{Constrained Array Test System}
\nomenclature{$CSV$}{Comma Separated Values}
\nomenclature{$Discretized$}{Process of transferring models and equation into discrete counterparts.}
\nomenclature{$FireEye$}{See ACTS}
\nomenclature{$MOLS$}{Mutually Orthogonal Latin squares}
\nomenclature{$MSI$}{Microsoft Installer}
\nomenclature{$NIST$}{National Institute of Standards and Technology}
\nomenclature{$NuSMV$}{Re-implementation and extension of SMV}
\nomenclature{$OATS$}{Orthogonal Array Testing Strategy}
\nomenclature{$PICT$}{Pairwise Independent Combinatorial Testing}
\nomenclature{$SMV$}{Symbolic Model Verifier}
\nomenclature{$WBPairwise$}{White-box Pairwise}
\printnomenclature
\end{IEEEkeywords}
\section{INTRODUCTION} \label{introduction}
Pair-wise testing is a type of combinatorial software testing, used in test case generation. Pair-wise testing helps prune the combinatorial explosion that can occur when attempting to test a system with many input options. Pair-wise testing often yields good results at a low cost. However, Pair-wise testing may miss 10-40\%+ of system bugs \cite{Go2016}. The set of values for each input is obtained from the component’s requirement \cite{Mathur201402}.  Pair-wise testing covers all possible pairs of parameter values by at least one test \cite{Go2016}.
\par
\begin{figure}
\centering
        \includegraphics[width=0.5\textwidth]{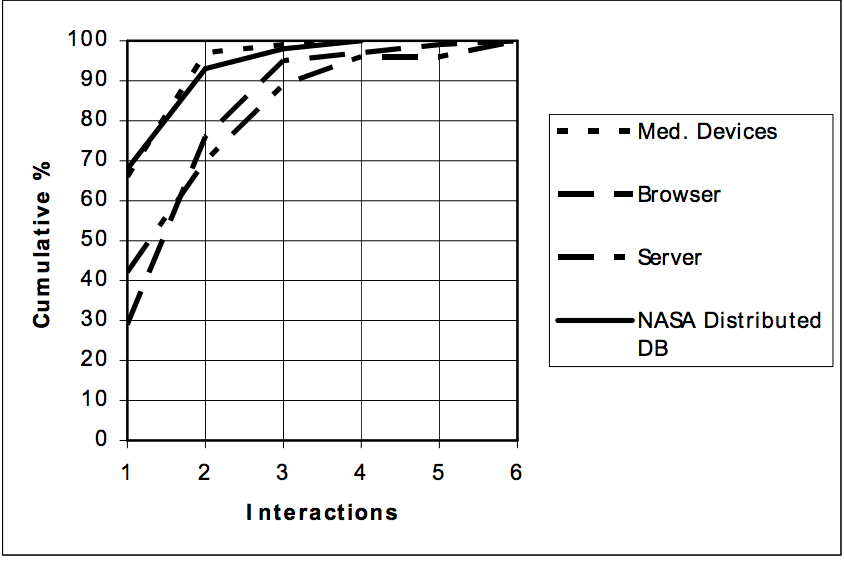}
     \caption{Error detection rates for interaction strengths 1 to 6 reprinted from NIST, PRACTICAL COMBINATORIAL TESTING, 2010}
\label{fig:errordetectionrates}
\end{figure}

There have been many studies on the efficiency of Pair-wise testing \cite{Go2016,cohen}. One such study showed, on average, 67\% of failures are caused by one parameter, 93\%  of failures are caused for 2-way pairs, and 98\% by 3-way combinations \cite{Go2016}. The results can be seen in \cref{fig:errordetectionrates}. Other studies such as \cite{cohen} agree with the findings of \cite{Go2016}.
\begin{quotation}Bellcore found that most field faults were caused by either incorrect single values or by an interaction of pairs of values. Our code coverage study also indicated that pair-wise coverage is sufficient for good code coverage \cite{cohen}\end{quotation}
This means that it is not necessary to attempt to test triples, quadruples, or more. The majority of defects will be caught and covered in the pair range. The goal of pair-wise testing is to generate minimal test case variants, while ensuring parameter combinations are covered. ``Testers can create effective and efficient test plans, often faster than by traditional methods entailing hand optimization \cite{cohen}.'' 
\par
There are many different names for pair-wise testing. You might encounter it being called pair-wise testing, 2-wise testing, All-pairs testing, OATS, t-wise testing, and possibly by several more names \cite{INDUCTIVEIntroductionTesting}. Despite the many names used to refer to pair-wise testing, typically, it is used in reference to 2-wise testing. A better term to encompass the generic model is t-wise (or t-way) as it does not refer to one particular degree of thoroughness. The thoroughness of pair-wise testing, called the strength or T, depends on the number of variables being processed. Testing thoroughness is measured in degrees from 1 to 6, with 1 being the lowest degree and 6 being the highest complexity \cite{INDUCTIVEIntroductionTesting}.
\par
The goal of pair-wise testing is to provide significant test coverage by testing every pair of options. Each pair of options will occur in, at least, one test case \cite{IntroductionToCombinatorial}, but may occur in more than one test case. Pair-wise testing creates a tabular representation of the program's factors. The rows of the table represent the combinations of options for each factor. There must be at least one test per row. Some options are fine by themselves; however, a defect may arise when one option is coupled with another option. Boundary and equivalency testing does not catch these faults and unwanted interaction occurs \cite{Watkins2010TestingProcess}. It is possible to test higher t-values of thoroughness; however, by increasing these test efforts, there are diminished benefits in terms of finding faults. The cost of covering those is disproportionately high and often unlucky to find bugs in these \cite{Watkins2010TestingProcess}. Pair-wise testing allows you to pack more coverage into fewer software tests, which also helps reduce the cost of testing.
\par
Pair-wise testing is a black-box test design technique. This means that pair-wise ignores the internal mechanisms and inner workings and instead focuses solely on the outputs. Pair-wise testing is used for testing unconstrained options (options are those that are independent of each other). Any options for any factor can coexist with any other option for any other factor. Configuration testing is a classic example of that. Pair-wise testing is used to control combinatorial explosions, related to testing unconstrained options. 
\par
Pairwise testing works well with agile development methodologies. Studies have been conducted and the results showed that as the thoroughness of pair-wise testing increases, the amount of faults found increases. According to Inductive no bug has ever been found that required more than a 6 degree of thoroughness \cite{INDUCTIVEIntroductionTesting}. As the thoroughness of pair-wise testing increases, so does the effort required to do the tests. Higher degrees of thoroughness typically require the aid of tools to assist in generating the test pairs and test cases. Luckily, there are an abundance of pair-wise/t-wise test generation tools available to assist in generating test pairs \cite{PairwisePairwiseTools}. A thoroughness of degree 2 (true pair-wise) is considered by many to be a good trade off and balance between thoroughness and effort. Often, pair-wise testing for lower degrees of thoroughness (1-2) can usually be done by hand if the number of factors are low.
\par
For testing mission or safety critical systems, pair-wise testing may be a good start, but other techniques should be used in conjunction to maximize test coverage. In the past, it has been difficult to use pair-wise testing for thoroughness above degree 3. In recent years there haven been new algorithms created to make combinatorial testing practical for higher values of T (4-6). One such tool is ACTS (previously FireEye an extension of IPO) created by the United States government at the National Institute of Standards and Technology \cite{NISTNISTACTS}. NIST is an agency of the U.S. Department of Commerce and was founded in 1901 and is the nation's oldest physical science lab \cite{NISTNISTACTS}.
\par
In pair-wise testing, each option and pair of options are represented about equally as a percentage of total configurations. In software testing, there are three types of fault modes; single, double, and multi. The simplest faults are single-mode faults. Single-mode faults occur when one option causes a problem regardless of the other settings. Double-mode faults are another type of defect that occurs when two options are combined. Multi-mode faults occur when three or more settings combine to produce the bug. 
\par
Complete coverage is usually not necessary. Most field faults are caused by either incorrect single values or by an interaction of pairs of values. The number of tests is, typically, $O(nm)$ where n and m are the number of possibilities for each of the two parameters with the most choices. The all-pairs method does not generate a balanced set of pairs and does not use orthogonal arrays.
\subsection{ALGORITHMS}
There are many different algorithms for approaching pair-wise testing, such as Orthogonal Array Test Strategy (OATS), Orthogonal Latin Squares, and Latin Squares, Mutually Orthogonal Latin squares (MOLS), coverage arrays, and mixed-strength covering arrays. These algorithms are different implementations of pair-wise test case pair generation. These algorithms and their implementations and use cases are outside the scope of this paper. There have been advances in these algorithms that allow for higher level thoroughness testing such as for 4-wise, 5-wise, and 6-wise testing.
\section{EXAMPLES}
\subsection{}
\begin{figure}
        \centering
                \i\includegraphics[width=0.5\textwidth]{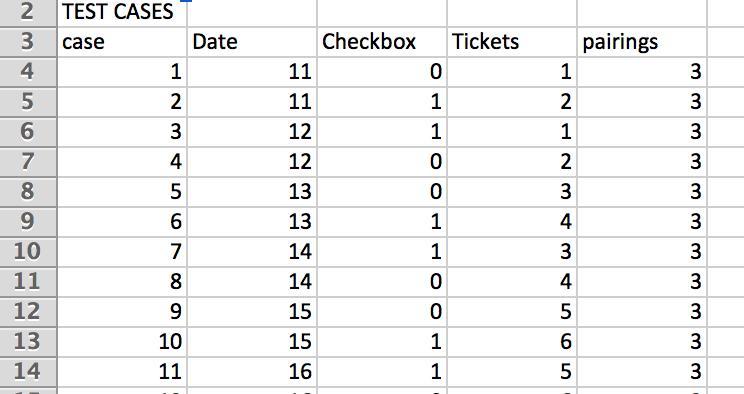}
        \caption{Sample of test cases generated from AllPairs.}
        \label{fig:testcases}
\end{figure}

Imagine a web based application that allows the user to purchase tickets to an event, similar to Ticketmaster. This application contains a date picker element that allows the user to choose 1 date out of 10 upcoming dates, a check-box to confirm user agreement to terms of service, and a select menu with the numbers 1-10 representing the number of tickets needed. A check-box is a Boolean representation as it can only be in one of two possible states: on or off. The total number of test cases can be calculated as \(10*2*10=200\). I installed the AllPairs Perl suite created by software tester, author, and former board director of the Association of Software Testing, James Bach \cite{BachTestInc.}. I used the AllPairs tool to generate the test cases and pairing data. Using this tool I reduced the number of test cases exactly in half to 100. An expert from the test case generation result from AllPairs can set can be seen in \cref{fig:testcases}.
\par
\subsection{}
Imagine you have three user inputs. The user can choose the type of car, the color of the car, and when they will purchase it. Possible car types are Ford or Mazda, colors are purple, blue, and silver, and purchase times are either dawn, day, or night. \Cref{table:example2wise} lists possible input values. An an exhaustive test would include ($3*2*3=18$) test cases for this small set of data. Using pair-wise testing and the 2-wise model we can reduce these 18 test cases to only 9 and still cover the same input scenarios as can be seen in \cref{table:example2wisetestcase}.
\begin{table}[ht!]
\centering
\begin{tabular}{||c c c ||} 
 \hline
Color & Make & Time \\
 \hline\hline
Purple & Mazda & Night \\
Blue & Ford & Day \\
Silver &  & Dawn \\
\end{tabular}
\caption{Allpairs example using 2-wise}
\label{table:example2wise}
\end{table}
\begin{table}[ht!]
\centering
\begin{tabular}{||c c c c c||} 
 \hline
case  &  colors & cars  &  times &  pairings \\
 \hline\hline
1  & purple & Mazda &  night &  3 \\
2 &  purple & Ford &   day & 3 \\
3  & blue  &  Ford  &  night  & 3 \\ 
4 &  blue  &  Mazda  & day & 3 \\ 
5  & silver & Mazda &  dawn &   3 \\
6  & silver &  Ford &   night &  2 \\
7  & purple & Ford &   dawn  &  2 \\
8  & blue &   \textasciitilde{}Mazda & dawn &    1 \\
9  & silver & \textasciitilde{}Mazda & day & 1 \\
\end{tabular}
\caption{Allpairs example test cases using 2-wise}
\label{table:example2wisetestcase}
\end{table}
\section{STRENGTHS}
Pair-wise testing is a fair compromise of cost and benefit \cite{Ramler2012CombinatorialImplications}. Pair-wise testing, typically, yields a significant reduction in the number of test cases that have to be ran without compromising functional coverage \cite{Ramler2012CombinatorialImplications}. Pair-wise testing lowers repetition, while maximizing variation. Pair-wise testing results in faults being found quickly and higher coverage of test inputs. The number of tests to be performed is reduced. Pair-wise testing generates small test sets, relative to the exhaustive test data set. Pair-wise tests rare conditions, produces high code coverage, finds faults faster, and lowers overall testing cost \cite{5197434}. Pair-wise works when there are more than 7 or 8 parameters and less than 300, depending on the interaction strength desired. Pair-wise is good when the processing involves interaction between parameters \cite{5197434}.
\section{WEAKNESSES}
The strengths of pair-wise testing fail when you do not properly select the right values to test. Pair-wise testing will not be effective if you choose the wrong input test data values. Pair-wise testing will not be effective if you do not have a good enough oracle. Defects are not always immediately observed and faults could occur behind the scenes without detection. High-risk combinations probably do not get enough attention. It's difficult to understand the n-wise connectivity within programs we are testing so its not clear that the pair-wise testing is an appropriate choice. Pair-wise does not consider the combinatorial characteristics of the system under test, can be very expensive at higher strength interactions, and may require high skill level in some cases (if formal models are being used) \cite{5197434}. Pair-wise is typically not useful when there is a relatively low number of parameters, when exhaustive testing is possible. Pair-wise is also not useful when there are not interactions between parameters \cite{5197434}.
\par
Pair-wise testing is not suitable for all data sets. Pair-wise testing allows data to be thrown away. This discarded data could be a missed important factor, and as a result, these important faults are not captured. When testing your software, there maybe be certain combinations of inputs that hold more weight or importance. In pair-wise testing, input values are not weighted. Instead, it is assumed that each pair of input values carries the same weight and significance on the output. \footnote{See \ref{PICT} for additional information}. 
\par
A software application may have many defects outside of the 2 parameter range. To reduce the risk of missing faults, testing using higher orders of thoroughness such as 3-wise, 4-wise, etc., should be considered.
\par
``Blindly applying Pairwise testing to combinatorial testing problems may increase the risk or delivering faulty software \cite{Bach2004}.'' In real-time or safety critical systems, any faults must be detected or loss of life may be a real possibility.
\section{TOOLS} \label{tools}
\begin{figure*}[tb]
\centering
\includegraphics[width=\textwidth]{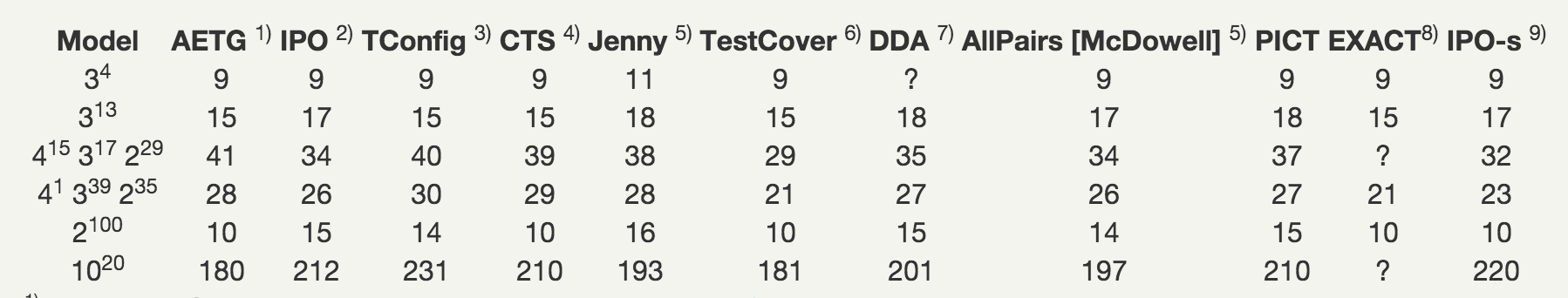}
\caption[caption]{Comparison of Efficiency, Source: \cite{PairwisePairwiseTools}\\\hspace{\textwidth}Number of test cases produced by different tools for the same model}
\label{fig:toolsComparison}
\end{figure*}

There are many free utilities for generating All-pairs tables such ass ACTS \cite{NISTNISTACTS}, AllPairs \cite{BachTestInc.}, PICT \cite{MicrosoftPairwiseScenarios}, Jenny \cite{JenkinsJenny:Tool}, Hexawise \cite{HexawiseHexawiseEasy}, and PWiseGen \cite{Flores2011PWiseGen:Algorithms}. \cite{PairwisePairwiseTools} lists 40 tools. \Cref{fig:toolsComparison} compares the efficiency of the top 9 tools for the same model.
\subsection{ACTS}
ACTS\footnote{Before February 2009 ACTS was called FireEye. You may encounter it referred with this name by papers referenced in this review.} is a test case generation tool created by NIST. ACTS is a standalone Java application that can be used from the command line or from the included GUI \cite{NISTNISTACTS}. ACTS can generate tests from 2-way to 6-way interactions. Occasionally, it is possible to generate some combinations that cannot actually be tested, because they do not exist (are not valid) for the systems being tested. ACTS offer the ability to specify constraints on the combinations generated, which allows you to specify invalid combinations. These tools allow you to generate sets of test configurations that do not include invalid combinations, but cover the essential combinations. These constraints can also be configured for combinations that are testable, but may be unnecessary, which can greatly reduce the overall number of tests \cite{NISTNISTACTS, Go2016}. ACTS can generate test cases that will cover 2-6-way combinations. The test cases can be exported in CSV, Excel spread sheet, as well as tab delimited formats \cite{NISTNISTACTS}. 
\subsection{Jenny}
Jenny is a tool for generating pair-wise tests. Jenny is written in C and has pre-built Windows binaries, available for download. Jenny is a public domain testing suite written by Bob Jenkins. Jenny can calculate N-wise test suites and defaults to pair-wise tests. Jenny is straightforward to use command line tool \cite{JenkinsJenny:Tool}. 
\subsection{Pairwise}
Pairwise is a Ruby-based tool for selecting a smaller number of test input combinations, using pair-wise generation rather than exhaustively testing all possible permutations \cite{WilkPairwiseRuby}. There is a Github repository that hosts the project \cite{WilkPairwiseRuby}. The last activity on the project was a few years ago; the project may be abandoned now. 
\subsection{AllPairs}
AllPairs is an open source test combinations generator, written in Perl by James Bach. It is released under the GPL 2.0 license. It allows one to create a set of tests, using pair-wise combinations methods, reducing several combinations of variables into a lesser set that covers most situations \cite{BachTestInc.}. It does not require any Perl modules to be installed other than those that ship with Perl. This was nice because I did not have to use CPAN to install any additional dependencies. The tool is fairly easy to use. After downloading and unzipping the source code, I generated the test cases found in \cref{fig:testcases} within a few minutes of familiarizing myself with the required format.
\subsection{Hexawise}
Hexawise is a Software as a Service provider that launched in 2009 and is based in Chapel Hill, North Carolina. Hexawise is a test case generator. Hexawise is in use at 3 out of the 10 largest IT consulting firms in the world, 3 of the 10 largest banks in the world, 4 of the 40 largest insurance companies in the world, and 2 of the 10 largest aerospace firms in the world. Hexawise boasts that is flexible and powerful, and can easily create more powerful software tests and optimized test generation. Hexawise allows you to generate complete test scripts. Hexawise allows you to maximize your testing thoroughness, eliminate gaps in testing coverage, and achieve greater coverage.
\subsection{PICT} \label{PICT}
PICT is a test case generation tool made by Microsoft \cite{MicrosoftPairwiseScenarios}. There is an MSI installer version provided by Microsoft that is available for download, but recently, PICT has become open source and has a git repository on the Microsoft Github account \cite{MicrosoftMicrosoftGithub}. PICT is written in C++ and can be cloned from the repository and built locally. The documentation for PICT seems thorough and includes instructions for building against multiple platforms. As of this writing, there seems to be movement on the project as there have been multiple commits pushed to the repository in the last few months. PICT allows you to force preference to certain values by attributing weight to the values. A weight can be any positive integer. PICT will use 1 if you do not specify a different weight explicitly. Weight values have no absolute meaning, and it is possible that the weights will be ignored. Look at the PICT documentation for a further explanation of how weights are used.
\subsection{PWiseGen}\label{PWiseGen}
PWiseGen is a test case generation tools and research platform. PWiseGen is highly configurable, extensible, and reusable \cite{Nguyen2012CombiningGeneration}. It also makes use of XML-based configurations to specify various parameters and new components implementing genetic operators. This tool is extensible because it makes use of interfaces to allow configurations and plug-and-play modules. Whenever someone creates a new module, only the underlying XML configuration file must change to make use of it \cite{Flores2011PWiseGen:Algorithms}.
\subsection{Notes}
When you run test case generation tools, you must run all tests that it identifies, you are not allowed to pick and choose. Test case generating tools assume that the order of the input variables is irrelevant. It is possible that you will need to test a dynamic user interface, where the order of value selections in controls is relevant. You might generate some test cases that are impossible to test. For instance if you have a list of operating systems and a list of browsers, you may get pairs such as Linux and Internet Explorer that are valid pairs, but are impossible to actually test. One possible solution to deal with these untestable pairs would be to use a tools like CATS created by AT\&T \cite{ATTTestcover.comBackground}. CATS helps testers pick just enough test cases. CATS analyzes cases from existing plans and suggests cases to add or remove. It can also generate a set of new test cases from test factor information. \cite{ATTTestcover.comBackground}
\section{PRACTICAL APPLICATIONS}
\subsection{A Case Study on Pairwise Testing Application}
The authors claimed that the success of pair-wise testing is based on the hypothesis that most defects are single-mode or double-mode defects. Pair-wise testing might not be necessarily better than ad hoc testing. They feel that "combinatorial testing is a shortcut, that is over promoted and poorly understood." The authors' goal was to compare the application of pair-wise testing with ad hoc testing applied to the same scenario. One specific scenario was selected for a target application.
\par
Input parameters were sorted, according to their conditional rules. Continuous input parameters were discretized through other testing techniques, such as boundary values and domain analysis testing. A complete combination of all input parameters would have generated an exhaustive 111,974,400 test cases. For the ad hoc testing analysis, a reduced set of test cases was initially generated by intuition of a specialist, based only on brute-force testing and experience in the business. For pair-wise, the All-pairs tool was used to generate 58 test cases. The generation of the ad hoc test cases oracle took roughly 4 times the effort to generate the pair-wise test cases oracle. Due to the large number of ad hoc test cases it was not practical to run them all, and instead only 14,041 were run.
\par
Evaluating the ad hoc test results took roughly 5 times the effort of the pair-wise test. The results showed that both test approaches could find the same defects on the software. The pair-wise test only took 19 test cases to find 10 defects, while the ad hoc test took 1,090 test cases. With much less effort, which represents costs, it was possible to find the same defects with pair-wise testing as was found by ad hoc testing. Other scenarios may be different, and such benefits may not be as obvious. The success of any test scenario highly relies on the testers' skills, experience, and knowledge on the business rules. With this in mind, focusing on quality as the main goal of the testing process, the pair-wise testing approach might help reach a satisfactory level of quality, by covering a more comprehensive field of test cases, with less effort spent, resulting in a more comfortable set of results for quality assurance and delivery \cite{MonteiroCassiano}. 
\subsection{White Box Pairwise Test Case Generation}
The authors of this paper presented a white-box extension to traditional black-box pair-wise test case generation. This extension selects additional test cases for the system, based on specifications for one or more internal sub-operations. They developed an algorithm for generating test cases for the full system, which achieve pair-wise coverage of the sub-operations. The authors based their WBPairwise algorithm on a case study for an elevator door control mechanism. The system has 14 parameters and served 3 floors, which yielded an exhaustive testing total of 2,359,296 test cases. They applied both the pair-wise method and their own WBPairwise to 500 different input parameter sets with different orders.  The results of their testing indicated that white box pair-wise testing is both practical and effective. WBPairwise alone performed nearly as well as pair-wise.  The authors concluded that the number of test cases generated and the algorithm execution run times are reasonable. They also could show that White box test sets are effective at revealing faults, and when combined with black-box tests such as pair-wise, they could improve the fault detection by nearly 4\% \cite{KimJangbok}.
\section{MODERN RESEARCH}
\subsection{Relationship between pair-wise and MC/DC testing: Initial experimental results}
Pair-wise testing is good at fault detection, but not for testing logical expressions. MC/DC was developed to test logical expressions. One draw back to MC/DC is that test generation can be a burden. The authors of this paper thought that it could be a good move to integrate the benefits of both MC/DC and pair-wise testing together. Due to the complexity of such a task, a smaller method was made to move forward.
\par
In this paper, the authors, Dr. Sergey Vilkomir and David Anderson, evaluate the level of MC/DC coverage for pair-wise test cases among different scenarios. These levels are then compared to the level of MC/DC coverage for random test cases. Tests of different size input variables, complexities, and logical expressions were used. Results showed the pair-wise test cases had a higher level of MC/DC coverage. These results were compared to the random test cases and supported the hypothesis \cite{Vilkomir}. Pairwise testing showed high levels of fault detecting, but failed to effectively detect faults in logical expressions. This failure led to the creation of MC/DC to test faults in logical expressions. Unfortunately, the test case generation can be time consuming and complicated. In this paper, the authors experimented to create a hybrid of pair-wise and MC/DC testing, which would integrate the benefits of both approaches.
\par
The authors stated it was important to evaluate the level of MC/DC coverage for pair-wise test cases in different situations and whether using pair-wise test cases as the basis for MC/DC testing provides benefits, compared to other techniques of test selection \cite{Vilkomir}. The authors divided their experiments into two different modes. Mode one focused on testing different number of input variables. Mode two focused on fixed input variable sizes. The results for both Mode 1 and Mode two showed that pair-wise coverage for MC/DC was higher than that of random testing. The results of Mode 1 showed that the average level of MC/DC coverage was 67.1\% for pair-wise and 62.5\% for random testing. The results from these experiments were deemed to be promising and the authors stated that it could yield future research pertaining to MC/DC coverage for mixed sets of logical expressions, MC/DC coverage of t-wise test cases, and MC/DC coverage for pair-wise and t-wise test cases for real world software applications, with the final goal to be creating a method that can extend pair-wise test cases to achieve MC/DC coverage closer to 100\% and a tool that can help in automating those methods. 
\subsection{Automated Combinatorial Test Methods: Beyond Pairwise Testing}
The authors felt that 2-way coverage, while being popular and low cost, was insufficient for mission-critical software \cite{Kuhn2008PracticalPairwise}. Combinatorial testing, except for pair-wise, is not used, because we lack good algorithms to perform complex combinations \cite{Go2016}. There are more tests that are required for combinations beyond 2-wise. Only a handful of tools are capable of generating complex combinations of strength 3 or 4. Some may take several days to generate all combinations. This is the reason that pair-wise testing has become accepted as a standard approach to combinatorial testing, as it is traceable. According to NIST, all failures can  be triggered by a maximum of 4-way to 6-way interactions, so there is no need to do higher strength orders.
\par
According to the publication, many test cases shouldn’t be a barrier if they can be generated with minimal human intervention. This automation should result in a reduction of cost. The authors proposed several new methods, including modeling the system and using covering arrays. They utilized system modeling using SMV in the NuSMV application. They modeled the system as a simple state machine and then evaluated the model and processed the results into test cases. Although the methods they used required human intervention and engineering judgment to define the formal model of the system, the automation of the test generation allowed them to be much more thorough than with other methods. This allowed them to test   all variable interactions to their specified strength and produce stronger mission critical software.
\subsection{PWiseGen: Generating test cases for pair-wise testing using genetic algorithms}
There is currently not known an efficient and optimal solution to find the smallest set of test cases. The time required to generate parameters for test cases grows exponentially. This paper formulates the problem of finding a pair-wise test set as a search problem and applies a genetic algorithm to solve it. Genetic algorithms are techniques that simulate the natural process of evolution \cite{Flores2011PWiseGen:Algorithms} which continues optimizing itself finding better solutions for problems. 
\par
The authors attempt to find a genetic algorithm, capable of minimizing the number of test cases that contain all pairs of input values to a software system to perform pair-wise testing. One issue of search-based/genetic algorithms and their applications to pair-wise testing is there are often too many variables to the algorithms. The large number of variables makes it difficult to tweak experiments. The authors experimented to test if a solution was available, using genetic algorithms for finding minimal test cases. They began by creating a genetic algorithm to represent test cases to serve as the chromosome to the problem.  This allows the genetic algorithm to iterate and crossover mutations to occur. Their work showed competitive results, compared with existing approaches and tools for pair-wise testing. The main contribution of the authors is a testing tool, called PWiseGen and released it as open source. PWiseGen is discussed in more detail in \cref{PWiseGen}. 
\section{CONCLUSION}
This paper presented a review of pair-wise testing, its strengths and weaknesses, tools to help with automated test case generation, and  use case reviews, and modern research. Without a doubt, pair-wise testing be considered a valid software testing strategy for generating test cases. However, this test strategy is only as good as the tester implementing it. The responsibility for good testing lies with the tester, not the tool being used. Pair-wise testing cannot replace an experienced tester, for example, knowing which input combinations are important and critical and making sure they are not excluded by the tool being used. Tools are meant to aid and make work easier, not replace the worker. When used properly, pair-wise test set generation is an important technique that can help you produce better software systems.
\par
As seen in \cite{Vilkomir} and \cite{Bach2004} pair-wise testing is not the magic solution to software testing. There isn't one testing solution that is best, nor is there a set of ``best practices'' that can be applied to any software component. The type of testing that you perform has to be determined by examining the data. Coupled with other software testing strategies, it is possible to gain high levels of coverage even on complex input systems \cite{BachTestInc.}. It is up to the software tester to know the application they are testing to determine which testing strategies and input values will yield the best results.
\par
Pairwise testing does not take risk into account since all pairs of inputs are treated equally. In real world applications, it is not usually the case that all pairs of inputs should be treated equally. In fact often you will find that in real world some combinations of inputs are more risky or prone to failure than others. It is very possible that test generation tools may not select the cases that are known to be high risk. It is our responsibility as testers to know and add these missing tests \cite{Go2016}.
\par
By itself, pair-wise testing will capture roughly 90\% of defects if utilized properly. For many software applications, this will be fine, as the cost associated with finding 100\% of all defects is not worth it. However, in safety critical systems pair-wise testing alone will not be enough to give assurance of the strength of the system. As seen in \cite{KimJangbok} and \cite{Vilkomir} it may be possible to couple traditional 2-wise testing models with other methods to gain full coverage.
\bibliography{Mendeley}
\bibliographystyle{IEEEtran}
\end{document}